# Exciton Energy Routing via Defect Networks in hBN/2D-Perovskite Hybrids

Sara Darbari[1,2,†*], Paul Bittorf[1,†], Leon Multerer[1], Fatemeh Chahshouri[1], Parsa Darman[1,2], Pavel Ruchka[3], Harald Giessen[3], Masoud Taleb[1,4], Yaser Abdi[1,5], Nahid Talebi[1,4*]

1 Institute of Experimental and Applied Physics, Kiel University, 24418 Kiel, Germany

2 Faculty of Electrical and Computer Engineering, Tarbiat Modares University, Tehran 1411713116, Iran

3 4th Physics Institute and Research Center SCoPE, University of Stuttgart, 70569 Stuttgart, Germany

4 Kiel Nano, Surface, and Interface Science KiNSIS, Kiel University, 24118 Kiel, Germany

5 Department of Physics, University of Tehran, Tehran 1439955961, Iran

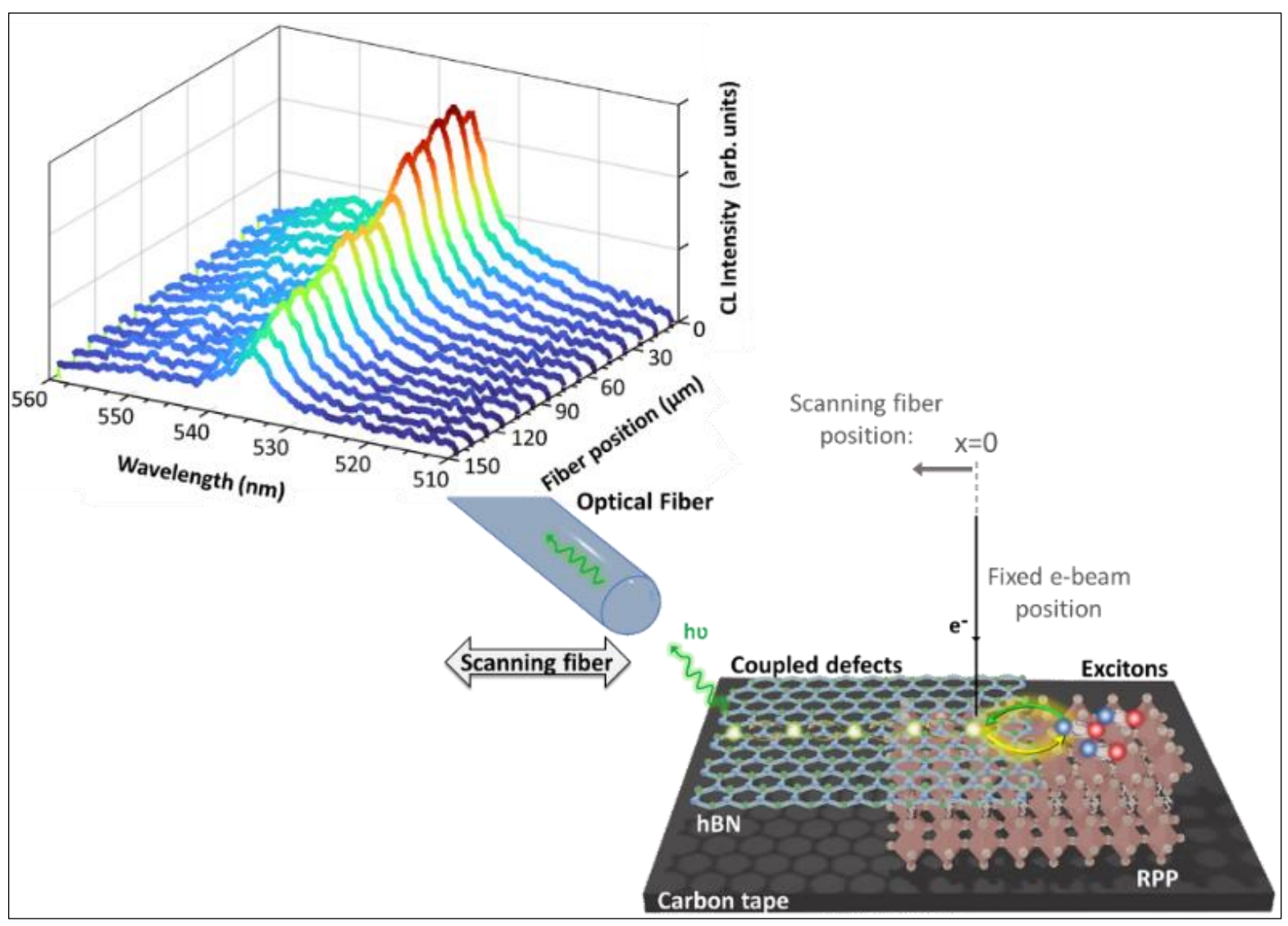

**ABSTRACT:** Excitons in two-dimensional Ruddlesden–Popper perovskites (RPPs) exhibit large and tunable binding energies, making them promising candidates for optoelectronic applications. In particular, long-range exciton energy transfer in these materials holds potential for light-harvesting technologies and nanoscale interconnects. Here, using cathodoluminescence spectroscopy, we demonstrate that exciton energy can be transferred over ultralong distances - up to 150 µm - in heterostructures composed of hexagonal boron nitride (hBN) and RPPs. This transfer is enabled by efficient exciton coupling to defect centers in hBN and subsequent defect-defect interactions. This mechanism not only facilitates long-range energy transfer, but also leads to enhanced luminescence intensity, narrower emission linewidths, extended exciton lifetimes, and reduced electron-beam-induced degradation. Owing to the high density of emitters within the hBN layers, the investigated van der Waals heterostructure emerges as a robust and stable hybrid platform. Our findings open promising pathways for room-temperature excitonic devices with enhanced performance, including quantum transducers, light-harvesting systems, and optoelectronic interconnects.

**Keywords:** 2D perovskite, cathodoluminescence, hBN/RPP heterostructure, exciton, defect, energy transfer

## Introduction

Two-dimensional (2D) perovskites, composed of a stack of 2D quantum-well layers and organic spacers, exhibit extraordinary optical properties, in addition to higher stability in comparison with their 3D counterparts. These properties have established them as a promising candidate for various optoelectronic and photonic applications[1–10]. Benefiting from the quantum-confinement within the quantum-well layers, 2D perovskites exhibit high exciton binding energies at room temperature[11], comparable to monolayer 2D transition metal dichalcogenides (TMDCs)[12–15]. This excitonic behavior of 2D perovskites make them a valuable platform to achieve strong light-matter interactions and polaritonic responses[16–20].

Exciton energy transfer mechanisms in nano-materials have been extensively explored through various mechanisms, including Förster resonant energy transfer[21–28], Dexter energy transfer[29–32], incoherent photon-mediated energy transfer[33,34], and coherent polariton formation[16,18,35–40]. Here, we demonstrate that there exists another distinguished mechanism, particularly in heterostructures of perovskite and hexagonal boron nitride (hBN); i.e., exciton energy transfer mediated through coupling of excitons to a network of localized defects in hBN.

HBN is one of the most stable and optically transparent encapsulating 2D materials with a large bandgap[41–43]. Moreover, hBN itself hosts multiple classes of phonon-assisted quantum-emitting defects[44–48]. A variety of these emitters demonstrate photon anti-bunching, which is a prerequisite for the use of single-photon emitters in quantum optical networks. In addition, radiation damage and degradation of RPPs caused by electron-beams in characterization techniques such as cathodoluminescence (CL) spectroscopy, can be reduced by hBN-assisted encapsulation[49–51].

Here, we realize heterostructures of mechanically exfoliated hBN and 2D Ruddleson-Popper perovskite (RPP), $(BA)_2PbI_4$, where BA is butyl ammonium and the number of $PbI_4$-octahedras monolayers is $n$=1. Utilizing CL spectroscopy as a powerful technique for exploring the photon statistics of the emissions from individual defects and clusters in several kinds of 2D materials as well as exploring the exciton physics of semiconductors[52–58], we study the optical response of the realized hBN/RPP heterostructures. Our results show an enhanced CL intensity, reduction in the linewidth broadening of the emission peak, increased decay time for the hBN/RPP heterostructures in comparison with RPPs, which is attributed to the efficient exciton-defect coupling at the interface of hBN and RPP. Significantly retarded degradation of RPPs is also achieved in hBN/RPP heterostructures during exposure to electron-beam, allowing reliable CL measurements at low current regime. Moreover, benefiting from a newly-developed 3D-printed fiber-based CL setup[59], we study the exciton energy transfer mechanisms in hBN/RPP heterostructures via changing the collecting-fiber distance from the electron-beam excitation location. This study reveals an ultra-long-range incoherent energy transfer – as in contrast with exciton polaritons[12,17] – through extruded parts of hBN in the hBN/RPP heterostructure over distances of about 150 μm, which is attributed to defect-defect coupling and photon recycling through the defect emitters in hBN. Our findings establish hBN/RPP heterostructures as hybrid van der Waals systems that exhibit enhanced luminescence and ultra-long-range exciton energy transfer. They are also promising candidates for room-temperature excitonic applications.

## Results and Discussions

**Cathodoluminescence Enhancement and Decay Rate of the Excitons in hBN/RPP Heterostructures.** To explore the optical properties of the hBN/RPP heterostructure, we first utilize a fiber-collected CL spectroscopy technique developed in our group[59], in which the heterostructure is excited by the electron-beam and the emitted radiation is collected by an integrated multimode optical fiber with a 3D-printed lens at its facet. The collected CL light is guided into the analyzing path, as depicted in Fig. 1a. In this configuration, the sample and the fiber are mounted onto individual three-axis piezo stages for precise in-situ positioning inside a scanning electron microscope (SEM) chamber. In the first measurement, the fiber focal point has been adjusted and fixed on the electron-beam impact position on the sample, while the sample stage is moved in the $xy$-plane to expose different sample parts to the electron-beam (Fig. 1a).

The CL spectra of different locations on an hBN/RPP heterostructure have been acquired by focusing the electron-beam excitation and the collecting fiber on the RPP flake (green circle), hBN edge on the RPP flake (orange circle), and the hBN/RPP heterostructure (blue circle) (see Fig. 1b). The acceleration voltage is 20 kV and current has been fixed at 140 pA in the low current regime, measured at the sample position using an implemented Faraday cup.

The measured CL spectrum exhibits a broad excitonic peak with the central wavelength of 530 nm and FWHM of 25 nm for the pure RPP flake (green spectrum in Fig. 1d corresponding to the green circle in Fig.1b), whereas a 5 nm redshift and 75% reduction in the bandwidth are observed for the electron-beam positioned at the hBN edge (orange spectrum in Fig. 1d corresponding to the electron position marked with orange circle in Fig. 1b). The CL peak is further enhanced and becomes narrower when the electron-beam traverses the heterostructure farther away from the edge (blue spectrum in Fig. 1d corresponding to the electron position marked by blue circle in Fig. 1b). The presented CL measurements and the corresponding observed enhancements in the excitonic peak of the hBN/RPP heterostructure are in analogy with the previously reported enhancements in hBN/TMDCs[60–68], where the observed phenomenon has been attributed to transfer of the excited electrons and holes in hBN into the 2D TMDC as the narrow bandgap material in the heterostructure[60,61]. Furthermore, in other studies the reduced linewidth broadening in hBN-encapsulated TMDCs has been attributed to the reduced surface roughness, reduced substrate-induced charge trapping, and protection from electron-beam irradiation in TMDCs because of encapsulating hBN flakes[65,67]. However, such mechanisms cannot explain the observations reported here, i.e., the long-range exciton energy transfer in the extruded parts of the hBN flake in the hBN/RPP heterostructure, as will be shown below. Here, we attribute the observed redshift and reduced bandwidth, dominantly to the phonon-assisted decay of the excited excitons in RPP to the localized and narrow bandwidth defect states in hBN, where the resonant energy of the latter is near to the RPP exciton wavelength. The proposed mechanism will be elaborated in Fig. 4c.

The lifetime of the excited carriers in hBN/RPP heterostructure is measured using a fiber-based Hanbury-Brown and Twiss (HBT) intensity interferometer, allowing us to acquire the second-order autocorrelation function $g_{\mathrm{CL}}^{(2)}(\tau)$ of the measured CL intensity signal $I_{\mathrm{CL}}(t)$ (Fig. 1c)[63,69–71]. This helps us to gain insight into the coherent and incoherent interaction processes of high-energy electrons within the heterostructures, where multiple quantum states are excited by each incoming electron. Since the excitation happens on extremely short interaction times of less than 1 fs, an inherent time synchronization is established between the emitters resulting in an observed bunching effect. These excitations diffuse and decay radiatively over time[72], hence, the measured second-order autocorrelation function ($g_{\mathrm{CL}}^{(2)}(\tau)$) displays a bunching effect at zero delay ($g_{\mathrm{CL}}^{(2)}(0) \gg 1$). Moreover, the decay rate of the bunching peak reveals the radiative lifetime of the ensembles of emitters. $g_{CL}^{(2)}(\tau)$ of the hBN/RPP heterostructure at different electron-beam excitation positions are measured (shown by colored circles in Fig. 1c), the results of which are plotted in Fig. 1e by the corresponding colors. For each sample position, the model function

$$g_{CL}^{(2)} = 1 + g_0 \cdot \exp\left(-|\tau|/\tau_d\right), \qquad (1)$$

was fitted to the experimental data[71]. $g_0$ represents the amplitude of $g_{CL}^{(2)}$ at the delay time of $\tau = 0$ and $\tau_d$ is the decay time of the excited state, or its lifetime. At all electron-beam excitation locations, the bunching peaks exhibit similar magnitudes of 32.5, 28, and 20.5 for the RPP (brown curve), hBN/RPP heterostructure (red curve), and the extruded part of hBN (orange curve), respectively. These values are significantly higher than the reported bunching peaks for hBN/TMDC heterostructures at a comparable current level[51]. However, the peak of $g_{CL}^{(2)}$ is the smallest when the electron-beam excites the extruded part of hBN (orange circle in Fig. 1c), away from the underlying edge of RPP in the hBN/RPP heterostructure. Moreover, the measured decay time for the extruded part of the hBN flake (51.2 ns) is longer than that for the hBN/RPP (49.7 ns), and the latter is longer than that for the pure RPP (47.2 ns). The observed increased lifetime in Fig. 1e is consistent with the previously observed decreased CL peak bandwidth for the hBN/RPP heterostructure in comparison with the RPP in Fig. 1d. It is notable that the measured longer decay time, implying on the slower annihilation of the excitons in the hBN-passivated RPP, can be attributed to the reduced surface roughness, surface passivation and protection against electron-beam irradiation by the top hBN flake[65,67].

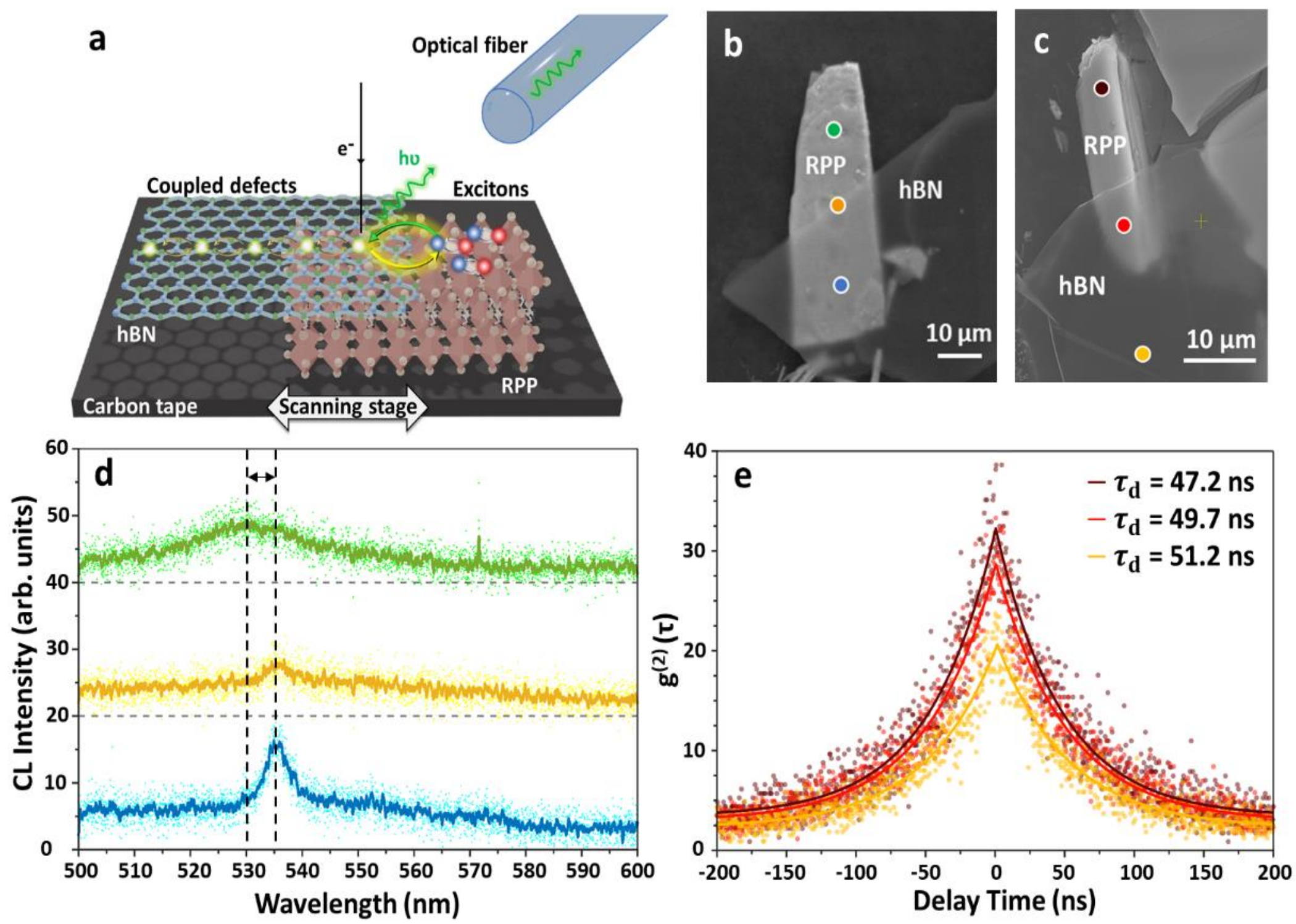

**Figure 1: Enhanced CL peak and life-time measurements facilitated by a fiber-assisted CL spectroscopy system.** (a) Schematic of the fiber-collected CL spectroscopy setup used for the hBN/RPP heterostructure. The emitted light is collected via a fiber with a 3D-printed lens on its facet and guided into the analyzing path. The focal point of the lens is fixed at the electron-beam incident point on the sample, and the sample lateral position can be changed by a scanning stage under the electron-beam. (b, c) SEM secondary electron images of the investigated hBN/RPP heterostructures, wherein colored circles show the electron impact positions. (d) Fiber-collected CL spectra for electron positions marked by the correspondingly colored circles in (b). (e) Second-order auto-correlation measurements ($g^{(2)}(\tau)$) from different electron impact positions, marked by the correspondingly colored circles in (c), revealing electron-beam-induced synchronization of emitters and photon bunching. Associated lifetimes of excitations at different areas of the sample are depicted as well.

**Long-Range Exciton Energy Transfer in hBN/RPP Heterostructures via Exciton-Defect-Defect Interactions.** Next, we investigate the hBN/RPP heterostructure by a mirror-collected CL spectroscopy system (Fig. 2a) at a low current level of 200 pA and acceleration voltage of 20 kV. In this measurement, a parabolic aluminum-coated mirror efficiently collects and collimate the emitted CL radiation and projects it into the analyzing path.

The sample is scanned by the electron-beam (blue and red arrows in Fig. 2b), while the collecting optics is remained at a fixed position. This way, higher spatial resolution is achieved, as compared to our previous results based on the fiber-collected setup where the stage was moved instead of the electron-beam. The measured energy-distance CL intensity maps for scanning the electron-beam along the blue and red arrows are displayed in Fig. 2c and Fig. 2d. Fig. 2c confirms the enhanced CL excitonic peak at the hBN/RPP heterostructure in comparison with the RPP, revealed by a step-like decrement in CL intensity as electron-beam passes the hBN edge over the RPP flake at the distance of about 2 µm. Moreover, while electron-beam passes over the edge of the underlying RPP flake at the distance of about 3 µm in Fig. 2d, the excitonic peak is significantly decreased, but keeps nearly a constant intensity on the extruded part of the top hBN flake up to distances of 15 µm. This observation is attributed to the excited excitons in RPP($n$=1) that are coupled to the nearly-matched electronic states of defects in hBN. Interestingly, a prominent CL signal is still observed when the electron-beam impinges the extruded part of the top hBN flake far away from the underlying RPP flake. In other words, Fig. 2d shows a long-range exciton energy transfer through hBN in the hBN/RPP heterostructure over a distance of several micrometers. Moreover, Fig. 2d reveals a spatially discretized profile for the CL emission peak in the extruded hBN after passing RPP edge along the red arrow. This phenomenon, together with the long-range exciton energy transfer are originated from the coupling of defects in hBN to the excitons in RPP at the hBN/RPP interface, as will be elucidated in more details below.

The acquired individual CL spectra in Fig. 2e reveal an excitonic peak at 520 nm for the RPP (position 'M' in Fig. 2b), an enhanced CL peak at 525 nm for the hBN/RPP($n$=1) heterostructure (position 'N' in Fig. 2b)[17,73], and a weak peak at 525 nm for the extruded part of hBN (position 'O' in Fig. 2b). Accordingly, the observed CL peaks and the slight redshift of about 5 nm (shown by red and orange dashed arrows in Fig. 2e) in the mirror-collected CL peaks of hBN/RPP heterostructure in comparison with RPP are in accordance with the results of the described fiber-collected CL apparatus in Fig. 1d. It should be mentioned that there exists a fixed wavelength offset of about 10 nm in the utilized fiber-collected setup with respect to the mirror-collected CL spectroscopy apparatus, revealing the agreement between the excitonic CL peaks of RPP and the hBN/RPP heterostructure. Additionally, a weak shoulder at around 570 nm is observed beside the main CL peak of the hBN/RPP($n$=1) heterostructure, which is attributed to the excitation of phonon sidebands of defects in hBN[44,47,74].

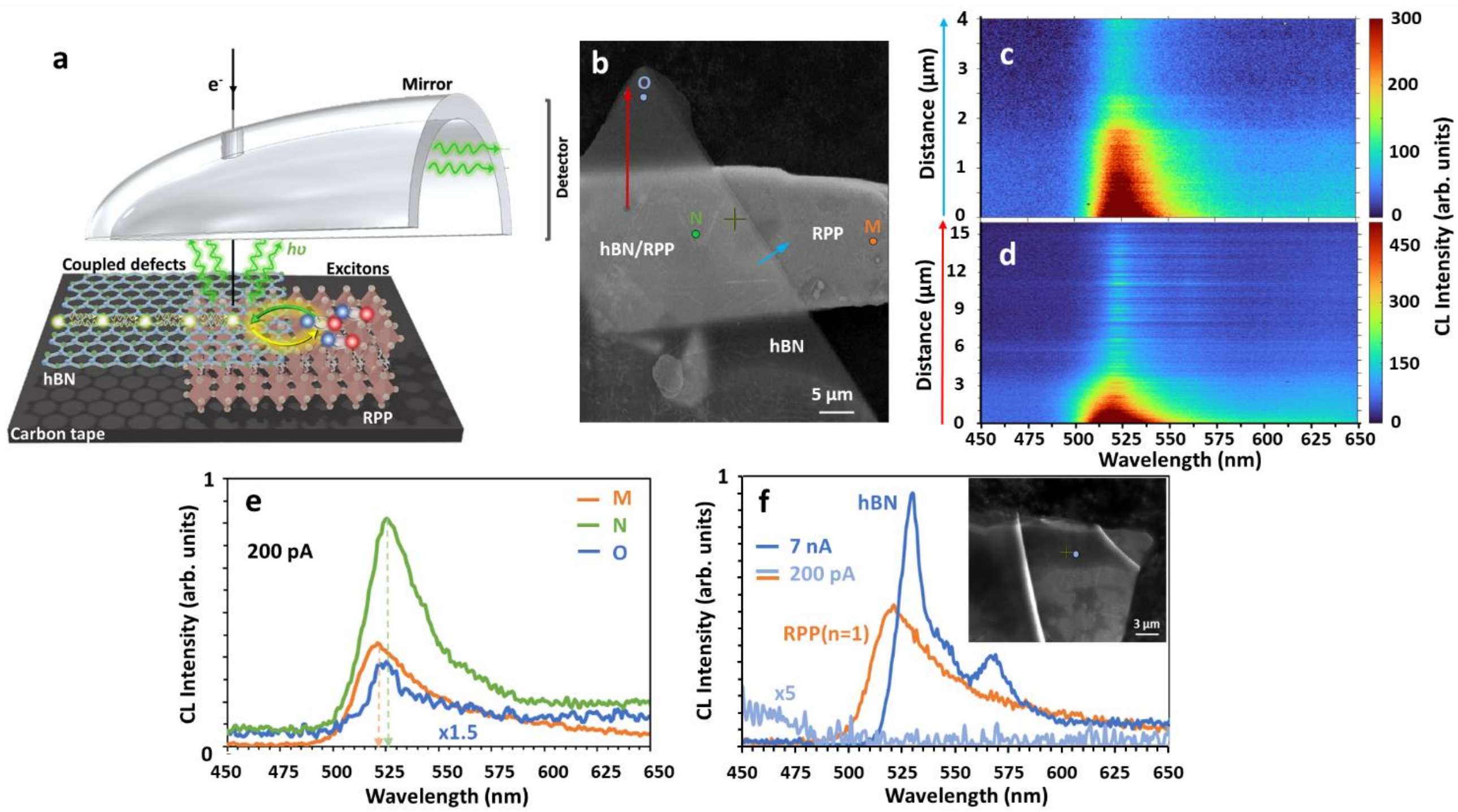

**Figure 2: Spatially-resolved cathodoluminescence response of the hBN/RPP heterostructure.** (a) Schematic demonstration of the mirror-collected CL spectroscopy setup used for the hBN/RPP heterostructure at current of 200 pA and acceleration voltage of 20 kV. The emitted light is collected by a parabolic mirror and directed toward a spectrometer. (b) SEM image of the investigated hBN/RPP heterostructure, wherein the blue and red arrows show the electron-beam scanning paths across the top hBN edge and the underlying RPP edge, respectively. (c), (d) The measured CL spectra along the blue (top) and red (bottom) scanning arrows in part (b). (e) Individual CL spectra corresponding to three different spots, marked as 'M', 'N' and 'O' in part (b). (f) CL spectra of a pure hBN flake (without RPP) integrated over a large area at the low current (200 pA) and high current (7 nA) regimes, superimposed on the CL spectrum of a pure RPP flake at low current (200 pA) regime. The inset displays the SEM image of the investigated pure hBN flake.

To elaborate on the proposed exciton-defect coupling in the hBN/RPP heterostructure, the CL spectrum of pristine hBN flake has been illustrated in addition to that of the pristine RPP flake (Fig. 2f). The CL spectrum of a pure hBN flake reveals no detectable CL peak at the low current regime (200 pA), similar to that current used for the RPP and hBN/RPP samples. However, benefiting from the inherent stability of hBN, we increased the beam current to a high current regime of 7 nA at the same acceleration voltage of 20 kV and acquired a clear CL trace of luminescent defects in the pure hBN with a strong and sharp peak at 530 nm and a weaker second peak at 570 nm, where the latter corresponds to the phonon side-bands. Fig. 2f reveals a significant spectral overlap between the excitonic peak at 520 nm in RPP($n$=1), and the luminescent defect peak at 530 nm in hBN, in addition to a weaker spectral overlap between the excitons and the second peak of hBN at 570 nm. This study solidifies the proposed efficient exciton-defect coupling, as the main reason behind the observed enhanced CL intensity, the slight redshift and the decreased bandwidth in the hBN/RPP heterostructure. Moreover, this study implies that the hBN defect energy states in this wavelength range are not excited efficiently by electron-beam at low current regime of 200 pA. However, coupling to excitons in hBN/RPP heterostructure enables a strong CL trace at 530 nm even at low current excitations, highlighting the high efficiency of defect excitations via excitons.

Linear plots of the CL intensity variations along the white arrow over the hBN/RPP heterostructure (inset), at peak wavelengths of 525 nm and 570 nm, better indicate the long-range energy transfer mechanisms (Fig. 3a). As observed, the enhanced excitonic peak intensity at 525 nm on the hBN/RPP heterostructure drops into a lower value by passing across the underlying RPP edge, but remains nearly constant over the rest of the scanning length of about 10 μm. A similar behavior is observed for the CL intensity at 570 nm; however, it has been significantly lower on the heterostructure and drops to comparatively lower intensities than that of the excitonic wavelength by passing across the underlying RPP edge. Furthermore, the intensity fluctuations demonstrated by the green curve in Fig. 3a confirms the closely located luminescent centers and the efficient coupling of excitons to defects in the hBN/RPP heterostructure.

CL hyperspectral images at the wavelengths of 525 nm, 570 nm (Fig. 3b), captured from the spatial region shown by the red dashed rectangle in the inset of Fig. 3a, better highlight the relevance of exciton-defect coupling scenario. The first and second columns in Fig. 3b display the CL intensity distributions spectrally filtered at selected wavelengths from 510 nm to 530 nm near the electronic level of defects, and at wavelengths of 560 nm to 580 nm attributed to the phonon side-band, respectively. At 515 nm, local bright spots are emerging across the whole scanning area, both on the hBN/RPP (left half of the frame) and on the extruded hBN (right half of the frame), which can be attributed to the efficient excitation of local defects in hBN coupled to the excitons. The observed luminescence and the emerged local bright spots show a highest intensity at around the excitonic wavelength of 520 nm, exhibiting lower intensities at shorter and longer wavelengths. The second column in Fig. 3b indicates lower CL intensities with less localized bright spots around 570 nm, following nearly the same spatial distribution around excitonic wavelength, strengthening the hypothesis of coupling between different luminescent centers associated with defects in hBN and the excitons in RPP.

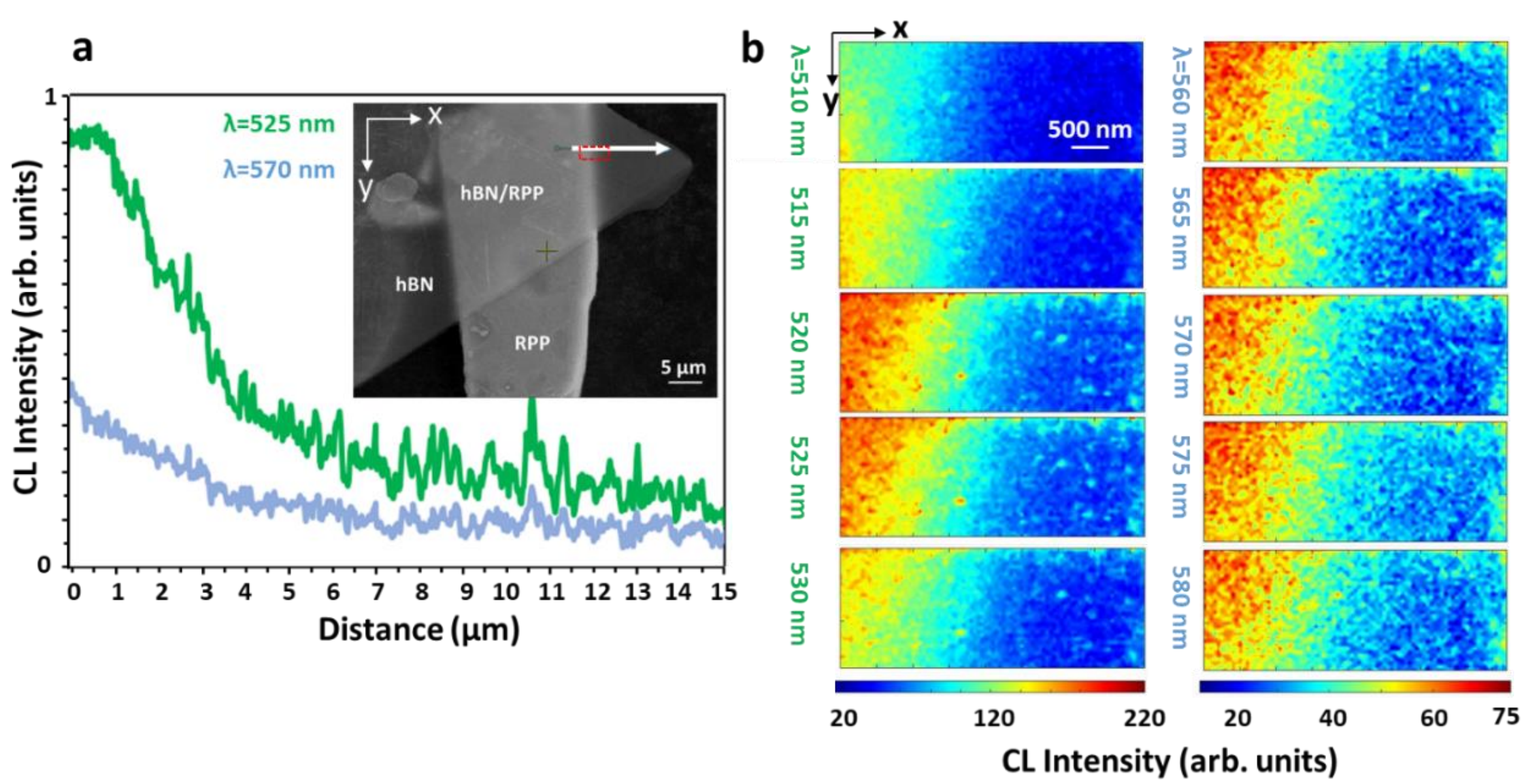

**Figure 3: Defects in hBN coupled to excitons.** (a) Intensity variations of two CL peaks in Fig. 2e versus the electron-beam scanning position, at 525 nm (green curve), and 570 nm (blue curve), along the path shown by the white arrow depicted in the inset, where the latter shows the secondary-electron SEM image the investigated hBN/RPP heterostructure. (b) Spatial profile of CL intensities at wavelengths around 525 nm and 570 nm, acquired over the dashed red rectangular box in the inset of panel (a).

Furthermore, we explore the long-range exciton energy transfer inside the hBN/RPP heterostructure by the fiber-collected CL spectroscopy technique, benefiting from a local detection scheme, due to the ability to individually move the CL-collecting fiber with a piezo stage, while the electron-beam impacts the sample at a fixed position. For this experiment, we investigate an hBN/RPP heterostructure with a large hBN flake positioned on top of an RPP flake, where the size of the latter is approximately 15×30 μm$^2$ (Fig. 4a). The beam current in this measurement is fixed at a low current regime (below 200 pA) and the acceleration voltage is fixed at 20 kV, similar to previous measurements.

During this measurement, the electron-beam position is fixed at the overlapping hBN/RPP heterostructure, marked by a white cross in Fig. 4a. First, the focal point of the fiber is placed and aligned at the electron-beam excitation spot, measuring the spectrum of the collected CL radiation around the excitonic wavelength of the hBN/RPP heterostructure, with a high spectral resolution. Next, the fiber's focal point is laterally moved away from the fixed electron excitation spot, while measuring the CL spectra along the scanning path of the fiber (shown by the white line in Fig. 4a).

The measured CL intensity (Fig. 4b), originating from the exciton-defect coupling, remains significantly strong even at a long distance of 150 μm away from the excitation spot on the hBN/RPP heterostructure, assuring an ultra-long-range exciton energy transfer through the hBN flake in the hBN/RPP heterostructure. This remarkably long-range excitonic energy transfer, in addition to the long lifetime of excited excitons (Fig. 1e) in hBN/RPP heterostructure highlight unique aspects of hBN/2D-perovskites heterostructures as an exciting class of optoelectronic materials in comparison with other platforms based on excitonic 2D materials at room temperature. Therefore, the hBN/RPP(*n*=1) heterostructures offer an ultra-long-range incoherent excitonic energy transfer, while benefiting from the inherently large exciton binding energy for RPP with *n*=1. In other words, the hBN/RPP heterostructure can significantly elevate the inherent trade-off paradigm in 2D RPPs between the direct exciton transport range that is minimized for *n*=1, growing by increasing *n*, and the exciton binding energy that is maximized for *n*=1, decreasing for higher *n* values[73].

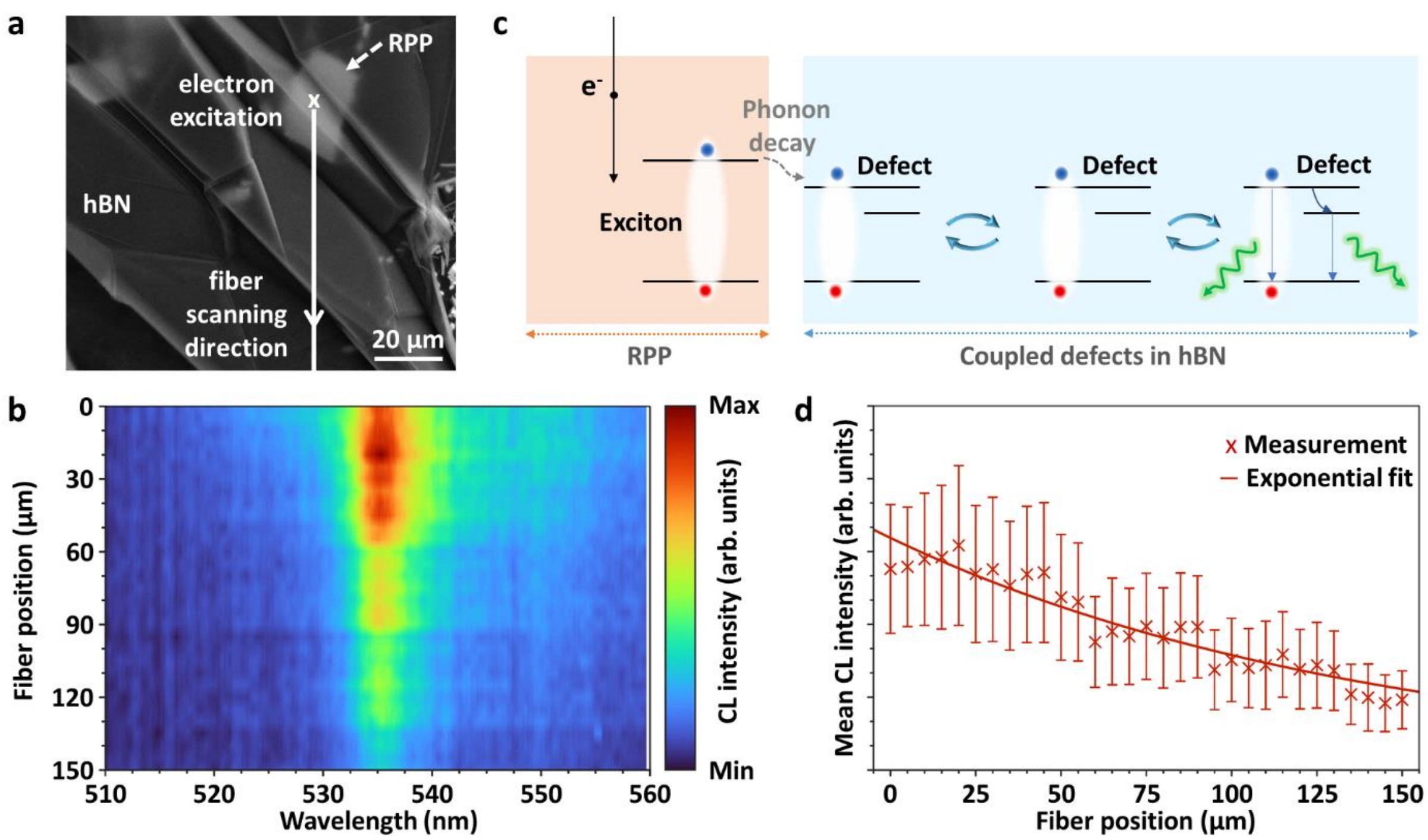

**Figure 4: Long-Range Exciton Transfer via Exciton-Defect Interactions.** (a) Secondary-electron SEM image of the hBN/RPP sample, investigated by the fiber-based CL spectroscopy method. The electron-beam incident point is fixed on the hBN/RPP heterostructure, shown by a white cross mark, while the fiber laterally scans the structure along the white arrow. (b) The measured CL spectra in the vicinity of excitonic wavelength, collected along the scanning path, which exhibits the long-range exciton energy transfer through hBN over a distance of 150 μm away from the hBN/RPP heterostructure. (c) The proposed physical principle of the observed long-range exciton energy transfer through coupled defects in the hBN/RPP heterostructure. (d) Attenuation of the measured excitonic peak intensity versus the distance from the electron impact position, fitted to an exponential curve with a decaying coefficient of $\gamma = 142.3\ \mu m$, which represents the exciton energy transfer length.

**Mechanism of the Incoherent Excitonic Energy Transfer in the hBN/RPP Heterostructure.** To explain the observed enhanced CL and the long-range excitonic energy transfer through the hBN/RPP heterostructure, here we elucidate the mechanism of the coupling between excitons of RPPs and defects of hBN (Fig. 4c). First, a moving high-energy electron (20 keV kinetic energy) excites an exciton in the underlying RPP flake, and the excited exciton couples to a defect center of about the same energy in the hBN flake via phonon decay. The excited defect further couples to the adjacent defect centers in the hBN layer via Coulomb interaction, partially transferring the excitonic energy through incoherent dipole-dipole coupling or defect-defect Förster energy transfer[25,26,28,75]. Defects may interact as well through radiation-assisted energy transfer or photon recycling, where the radiation of one excited defect can be reabsorbed by the adjacent defect subsequently. Photon recycling, enhanced by the reflection of the photons from the boundaries of the layers, particularly can play a significant role in the radiation-assisted energy transfer[33,34,76,77]. Defect-defect coupling, in addition to a sequence of photon recycling through defect emitters are believed to play significant roles in the observed ultra-long-range incoherent excitonic energy transfer in the hBN/RPP heterostructure. The elucidated phonon-mediated exciton-defect coupling explains the previously observed slight red-shift (about 5 nm) in the CL peak of the hBN/RPP heterostructure in comparison with the RPP excitonic peak (Fig. 1d and Fig. 2e). Due to presence of defect levels in hBN and their luminescence in this wavelength range (Fig. 2f), the proposed exciton-defect-defect coupling can lead to radiative recombination at the red-shifted excitonic wavelength in the hBN/RPP heterostructure (a schematic of the many-body interactions is shown in Fig. 4c). Furthermore, to quantify the effective length of this exciton energy transfer, we plot the measured excitonic CL peak intensity versus the scanning distance of the collecting fiber with respect to the electron impact position on the hBN/RPP heterostructure (Fig. 4d). By fitting an exponential curve ($\alpha \exp(-x/\gamma)$) to the acquired data set, as shown in Fig. 4d, where $\gamma$ represents the effective exciton energy transfer length equal to 142.3 μm for the investigated hBN/RPP heterostructure. This observation confirms the efficient and ultra-long-range exciton energy transfer through the coupled exciton-defect-defect system, originating from inherently large density of point defects in hBN with an efficient Coulomb correlation and extensive hoping. This behavior is further confirmed by measurements performed on other investigated hBN/RPP heterostructures (See Fig. 5, Supplementary Note 1 and Supplementary Figure 1).

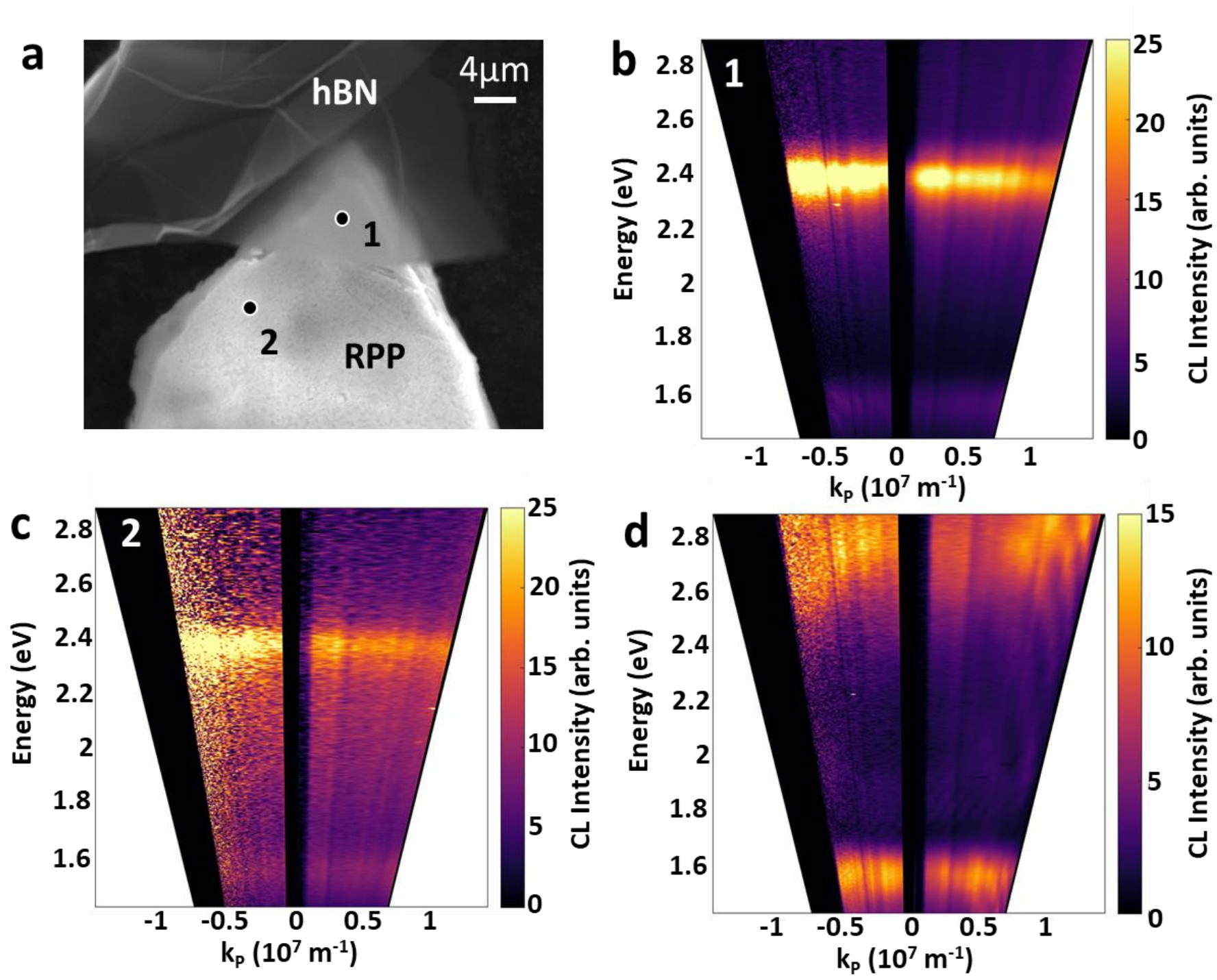

**Figure 5: The Energy-momentum map of the hBN/RPP heterostructures.** (a) The secondary electron SEM image of the explored hBN/RPP structure. The energy-momentum map of the hBN/RPP heterostructures acquired at position 1 (b) and position 2 (c). The observed nearly flat band at the excitonic energy for the hBN/RPP heterostructure rejects polaritonic or waveguiding behavior and is in accordance with the proposed mechanism of coupled local defects. (d) Energy-momentum map of the pure hBN flake explored in Fig. 2(f), provided for the sake of comparison. Electron-beam current and acceleration voltage have been 200 pA and 20 kV in all parts.

The long-range energy transfer mechanism observed in the hBN/RPP heterostructure is not due to the coupling of the exciton emission to the waveguiding modes of hBN flakes. This claim is substantiated by the fact that no coherent response, such as the generally observed Fabry-Pérot interference patterns or interferences observed due to the interaction of waveguiding modes with the edges of the flakes are observed here in the CL hyperspectral images[12,17], but rather randomly positioned emission centers are confirmed. Moreover, using the mirror-collected CL spectroscopy, the energy-momentum maps of the emitted CL are measured on the hBN/RPP($n$=1) heterostructure and the RPP flake (labeled as '1' and '2' in Fig. 5a), revealing nearly flat bands at the exciton energy of 2.4 eV (Fig. 5b, c). The presented flat dispersion (Fig. 5b) at exciton energy further validates a slow energy transfer mechanism as a result of exciton-assisted efficient excitation of coupled luminescent local defects in the hBN/RPP heterostructure, as opposed to the coupling of excitons with the waveguiding modes. The observed excitonic flat band aligns with the proposed energy transfer mechanism based on exciton-defect-defect coupling in the hBN/RPP heterostructure. For comparison, the energy-momentum map of the pure hBN flake is also provided (Fig. 5d). All energy-momentum maps have been measured at the low current regime of 200 pA and an acceleration voltage of 20 kV.

To rule out the artifacts related to the utilized fiber collection geometry and secondary electron excitation, we conducted a control experiment using a pristine RPP flake without an hBN overlayer (see Supplementary Figure 2). In this configuration, the electron-beam with a low current (<200 pA) was fixed on the RPP flake while the fiber collector was gradually displaced from the excitation point (Supplementary Fig. 2a). The resulting signal exhibited a short-range decaying behavior (part b) with around fifteenfold shorter propagation length, extending up to $\gamma$=9.6 μm (Supplementary Fig. 2c). Notably, this energy transfer range still exceeds the known exciton direct diffusion length in pure RPP—which is in the sub-micrometer regime—and thus lies beyond the spatial resolution of our fiber-based CL detection. Here, the measured spatial decaying coefficient can be dominated by effects of the secondary electron excitation and the finite point spread function of the fiber, with respect to the direct exciton transport range in pure RPP. Hence, the presented fiber-scanning CL measurement is not suitable for evaluating the short-range diffusion behavior, as opposed to the presented long-range energy transfer mechanism in the hBN/RPP heterostructure.

A mechanism contributing to the observed enhanced CL response in hBN/RPP heterostructure can be the secondary or back-scattered electrons from the underlying RPP flake that increase the emission yield from the defects in hBN. This behavior will be better clarified by the Monte Carlo simulations (See Supplementary Note 2 and Supplementary Figure 3). However, the major contribution is due to the multilevel defects in hBN and the exciton-defect-defect coupling in the hBN/RPP heterostructure, as discussed earlier and is solidified more in Supplementary Figures 4 and 5. A further mechanism that could play a role in the observed CL response in the extruded hBN flake is the multiple and sequential elastic and inelastic scattering events of the excited secondary and back-scattered electrons in the multilayer hBN/RPP structure. Generally, such mechanisms lead to a CL response from excitons, when the electron-beam excites the hBN flake up to approximately several 100 nanometers away from the underlying RPP edge. As shown by the Monte Carlo simulations by considering the multilayered geometry, the CL response approaches zero after 2 μm away from the edge of the underlying RPP flake (See Supplementary Note 2 and Supplementary Figure 3). Therefore, this mechanism is only relevant at short distances away from the edge of the RPP flake and cannot substantiate the observed ultra-long-range exciton energy transfer in hBN/RPP heterostructure.

Degradation effects in RPP are due to different factors such as the environmental humidity, and exposure to light or electron-beam[43,50,51]. After one month aging, the RPP flakes do not sustain a strong excitonic response, and the corresponding exciton-defect coupling and the enhanced CL peak at exciton energy are significantly faded, as well as the consequent long-range exciton energy transfer mechanism (see Supplementary Note 3 and Supplementary Figure 4). This observation clarifies the key role of excitons of RPPs and the resulting exciton-defect coupling in the observed enhanced and long-range CL behavior of the hBN/RPP heterostructure, ruling out a considerable contribution of the reflected electrons from the underlying RPP.

To elaborate more on the exciton-defect coupling in hBN/RPP heterostructure, hBN/RPP($n$=2) heterostructure are realized and probed by the mirror-collected CL spectroscopy (see Supplementary Note 4 and Supplementary Figure 5) at beam current of 200 pA and acceleration voltage of 20 kV. The measured energy-distance CL intensity maps and the individual CL spectra at different positions of the hBN/RPP($n$=2) heterostructure reveal a dominant CL peak at 580 nm and a weak peak at 530 nm on the hBN/RPP($n$=2) heterostructure. Considering the significant spectral overlap and the consequent coupling of the excitons in the pure RPP($n$=2)[73] at around 590 nm with the phonon side-band peak of defects in pure hBN at 570 nm, this experiment strongly confirms the dominant role of RPP's excitons in emergence of the enhanced CL peak in the heterostructure, which is observed at about 580 nm in hBN/RPP($n$=2). Moreover, the weak spectral overlap of the RPP($n$=2) excitons with the inter-band transition of hBN's defect-states, has also led to the emergence of a weak exciton-defect coupling peak at 530 nm in the hBN/RPP($n$=2) heterostructure. Therefore, it is proved that by changing the exciton energy of RPP in the hBN/RPP heterostructure, the energy of the coupled exciton-defect dominant peak is shifted from 525 nm for hBN/RPP($n$=1) to about 580 nm for hBN/RPP($n$=2). This experiment solidifies the mechanism of efficient exciton-defect coupling in hBN/RPP heterostructures, which can be tuned by changing $n$ in RPP.

## Conclusion

Benefiting from a high exciton binding energy at the room temperature, RPPs also exhibit a slow exciton annihilation rate and relatively long exciton diffusion lengths, compared to their TMDC counterparts. The combination of these extraordinary optical properties in addition to their simple synthesis and composition tunability entitles 2D RPPs as superior candidates for optoelectronic applications, while their inherent degradation in exposure to light and electron-beam have been remained as their main drawback.

Here, we have explored hBN/RPP heterostructures by CL spectroscopy, which has been allowed by the surpassed electron-beam-induced degradation of RPPs at low current regime. The hBN/RPP($n$=1) heterostructure have revealed enhanced luminescence intensity, narrowed emission bandwidth, increased decay time of the excited excitons, and a slight redshift in the excitonic wavelength of RPPs. Furthermore, benefiting from the dual scanning functionality of a unique fiber-collected CL spectroscopy setup, an ultra-long-range exciton energy transfer length of about 150 μm has been proved through hBN in the hBN/RPP($n$=1) heterostructure. The observed enhanced CL spectrum and the ultra-long-range exciton energy transfer have been attributed to the exciton-defect coupling in the hBN/RPP heterostructure, and the defect-defect coupling in addition to the photon recycling through the defect emitters in hBN. Realizing high-quality hBN/RPP($n$=1) heterostructures has intensively enhanced the exciton energy transfer, while keeping the benefit of the inherently highest exciton binding energy with respect to other RPPs with $n$>1, emphasizing the potential of hBN/RPP($n$=1) heterostructures for applications in active optoelectronics and light-harvesting devices.

## Methods

**Sample preparation**. To achieve 2D RPP($n$=1) bulk crystals[1], initially n-butylammonium iodide (BAI) is synthesized by gradually adding 25 ml 57% w/w aqueous hydriodic acid (HI) to 5 ml n-butylamine (BA) within an ice bath, while keeping stirring for 4 hours. Next, 500 mg of PbO powder is dissolved in a mixture of HI (57%,3 ml) and $H_3PO_2$ (50%,850 µl) solution at the temperature of 120°C, while stirring the solution for about 5 minutes until converting into a bright yellow solution of $PbI_2$. Thereafter, 1.5 ml of the synthesized n-$CH_3(CH_2)_3NH_3I$ (BAI) is added to the prepared $PbI_2$ solution, resulting in a transient black precipitation that is dissolved by keeping the solution temperature at 120°C. After 5 minutes in this condition, stirring and heating is stopped, letting the solution to cool down to room temperature, during which the orange bulk crystalline $BA_2PbI_4$ are emerged among the solution. Using vacuum filtering, $BA_2PbI_4$ products are separated and dried during a few days.

Next, to realize the van der Waals hBN/RPP heterostructure, the synthesized $BA_2PbI_4$ crystalline flakes have been mechanically exfoliated on to a double-sided adhesive carbon tape. For this purpose, a $BA_2PbI_4$ flake was picked up by a piece of scotch tape and after folding and unfolding for several times several flakes with reduced thicknesses were achieved, which were stamped on the carbon tape subsequently. HBN crystals have been purchased from the HQ Graphene Company. Utilizing mechanical exfoliation, thinned hBN flakes have been transferred on the RPP flakes that were previously transferred on the carbon tape. Overlapping of the large hBN flakes on the underlying RPP flakes, van der Waals hBN/RPP heterostructures are achieved.

**Degradation issue in cathodoluminescence spectroscopy of hBN/RPP heterostructures: Electron-beam-induced degradation in RPPs.** The presence of hBN flake on top of RPP significantly protects the RPP from electron-beam-induced degradation. In addition, to investigate the suitable measurement conditions for CL spectroscopy on RPP flakes, we systematically investigated the electron-beam conditions to find optimum settings. Exposing partially covered RPP flakes by hBN to high-current electron-beams reveals drastic degradation in the uncovered parts of the RPP flake. In contrast, the hBN cover results in efficient protection against electron-beam irradiation (Supplementary Note 5 and Supplementary Figures 6). Moreover, an acceleration voltage of 20 kV and a low current regime of about 200 pA is best used to efficiently excite excitons in RPP flakes, while avoiding the electron-beam-induced degradation, even without hBN protection (Supplementary Figure 7).

**Fiber-collected cathodoluminescence spectroscopy and time-correlated single photon counting**. The measurements shown in figures 1, 4, S1, and S2 have been performed by using a Thermo Fisher Quattro S field-emission scanning electron microscope (FE-SEM). Throughout the experiments, the acceleration voltage and the current of the electron-beam have been adjusted at 20 kV and 140 pA. A built-in amperemeter is connected to the sample stage in the SEM, measuring the effective beam current at the investigated specimen. Inside the SEM, a system of three perpendicular linear piezo-driven stages from SmarAct is mounted, holding an optical multimode fiber which can be moved independently from the sample stage along three axial degrees of freedom. The stages feature dynamic ranges of 83 mm along the x-axis, and 35 mm along the y- and z-axis with an accuracy of 1 nm in step size, allowing for a precise and high-spatial selectivity in the collection of the cathodoluminescence radiation by the fiber. For the experiments we used the optical multimode fiber FG400AEA from Thorlabs, which has a silica core diameter of $(400 \pm 8)$ µm with a broad spectral operation bandwidth from 180 nm up to 1200 nm, and a numerical aperture of $0.22 \pm 0.02$ for the flat fiber cross section. To improve the numerical aperture, and therefore the collection efficiency, a lens was 3D printed onto the face of the multimode fiber, using the Quantum X 3D printer from Nanoscribe GmbH [78–81]. The numerical aperture of the fiber with this micro-optic has been increased to 0.36 with a working distance of 400 µm approximately, while the fiber has a fixed inclination angle of 35° with respect to the sample stage. The electron-beam position and its spot size are controlled via the electron optics of the SEM. The presented fiber-collected CL configuration allows for adjusting the position of the focal point of the fiber with respect to the electron-beam impact position on the sample. The lateral (in-plane) position of the fiber is measured by using the secondary electron detector of the SEM apparatus, whereas the vertical (out-of-plane) position of the fiber is fixed at the position with the highest collected CL intensity from the investigated sample.

The fiber cable guides the collected CL radiation towards either a spectrometer or photomultiplier tubes. For the spectroscopy measurements we use the Teledyne Princeton Instruments HRS-500 spectrograph attached with the PyLoN 100BRX liquid-nitrogen-cooled CCD camera. It is equipped with three gratings which all have a groove density of 1200 $mm^{-1}$ with a spectral resolution of 0.1 nm and blaze wavelengths of 300 nm, 500 nm, and 750 nm, respectively. In addition, a Hanbury Brown-Twiss intensity interferometer has been utilized for performing time-correlated single photon counting and measuring the second-order autocorrelation function $g^{(2)}_{CL}(\tau)$ of the emitted CL intensity signal. This assembly consists of two PMA Hybrid 50 photomultiplier tubes from PicoQuant, the quTAG time-to-digital converter from qutools and the TT400R5F1B 1x2 multimode fiber coupler from Thorlabs with a 50:50 splitting ratio.

**Mirror-collected cathodoluminescence spectroscopy**. As a complementary measurement we also performed CL spectroscopy using an aluminum-coated parabolic mirror to collect the emitted CL radiation, which are shown in figures 2, 3, 5, S4, S5, S6 and S7. These measurements were performed inside a ZEISS Sigma field-emission scanning electron microscope, equipped with the Delmic SPARC CL system. Throughout these measurements the acceleration voltage of the electron-beam has been set to 20 kV, whereas the beam current has been kept at 200 pA to hinder electron-beam-induced degradation in RPPs, unless otherwise specified. The CL detector consists of an off-axis paraboloid mirror which is positioned with an accuracy less than 1 µm above the specimen and redirects the collected CL radiation onto the analyzing path and a CCD camera (Andor i-KonM) for further analysis. The parabolic mirror has an acceptance angle of $1.49\pi$ sr (NA = 0.97) and a focal distance of 0.5 mm. The exciting electron-beam can pass the mirror through a hole with a diameter of 600 μm located above the focal point. Utilizing a dispersive reflection grating and on occasion directing the CL radiation through a one-dimensional slit opening, the CL detector is capable of performing hyperspectral and energy-momentum imaging. The acquisition time for each pixel was set to 200 ms for hyperspectral imaging and 120 s for energy-momentum mapping.

## ASSOCIATED CONTENT

### Supporting Information

The Supporting Information is available free of charge on the ACS Publications website.

Supplementary Note 1: Complementary fiber-collecting CL measurements to confirm the long-range energy transfer in hBN/RPP($n$=1) heterostructures.

Supplementary Note 2: Monte Carlo Simulations of the scattered electrons in the hBN/RPP heterostructure.

Supplementary Note 3: The effect of RPP degradation on the optical response of the hBN/RPP heterostructure.

Supplementary Note 4: Exploring the hBN/RPP($n$=2) heterostructure by mirror-collecting cathodoluminescence.

Supplementary Note 5: Exploring the electron-beam-induced degradation in RPPs.

**Data availability**: The dataset generated during and/or analyzed during the current study are available from the corresponding author on reasonable request.

**Code availability**: The numerical code used to simulate the data are available from the corresponding author on reasonable request.

## AUTHOR INFORMATION

### Corresponding Authors

*Email: talebi@physik.uni-kiel.de
*Email: s.darbari@modares.ac.ir

### Author Contributions

S.D. and P.B. contributed the same to the paper. S.D. led the conceptual development, prepared the investigated samples, partially contributed to the measurements, analyzed and interpreted the results, and prepared the original draft together with N.T. P.B. organized the fiber-collected cathodoluminescence spectroscopy setup, performed the cathodoluminescence and time-correlated single photon counting measurements, interpreted the results, and collaborated in preparing the original draft. L.M. carried out the mirror-collected cathodoluminescence spectroscopy. F.C. conducted Monte Carlo simulations, interpreted the results and collaborated in preparing the original draft. P.D. synthesized 2D-perovskites and supported sample preparation. P.R. and H.G. supported the Fiber-collected cathodoluminescence spectroscopy, and edited the manuscript. M.T. and Y.A. conducted and supported the mirror-collected cathodoluminescence spectroscopy. N.T. acquired the funding resources, led the conceptual development, supervised the measurements, interpreted the results, and contributed to the original draft. All authors have reviewed and given approval to the final version of the manuscript.

† These authors contributed equally to this work.

## ACKNOWLEDGMENT

This project has received funding from the Volkswagen Foundation (Momentum Grant), European Research Council (ERC) under the European Union's Horizon 2020 research and innovation program under grant agreement no. 802130 (Kiel, ERC Starting Grant NanoBeam), grant agreement no. 101157312 (ERC PoC Grant UltraCoherentCL), grant agreement no. 101170341 (Kiel, ERC Consolidator Grant UltraSpecT) and grant agreement no. 101017720 (EBEAM), Alexander von Humboldt Foundation, from Deutsche Forschungsgemeinschaft (GRK2642), BMBF (Integrated 3DPrint, QR.X as well as QR.N), and EU (IV-Lab – 101115545).

## Supporting Information

# Exciton Energy Routing via Defect Networks in hBN/2D-Perovskite Hybrids

Sara Darbari[1,2,†*], Paul Bittorf[1,†], Leon Multerer[1], Fatemeh Chahshouri[1], Parsa Darman[1,2], Pavel Ruchka[3], Harald Giessen[3], Masoud Taleb[1,4], Yaser Abdi[1,5], Nahid Talebi[1,4*]

[1]Institute of Experimental and Applied Physics, Kiel University, 24418 Kiel, Germany

[2]Faculty of Electrical and Computer Engineering, Tarbiat Modares University, Tehran 1411713116, Iran

[3]4th Physics Institute and Research Center SCoPE, University of Stuttgart, 70569 Stuttgart, Germany

[4]Kiel Nano, Surface, and Interface Science KiNSIS, Kiel University, 24118 Kiel, Germany

[5]Physics Department, University of Tehran, Tehran 1439955961, Iran

† These authors contributed equally to this work.

*Corresponding Authors:*

**Email*: talebi@physik.uni-kiel.de

**Email*: s.darbari@modares.ac.ir

## Supplementary Note 1: Complementary fiber-collecting CL measurements to confirm the long-range energy transfer in hBN/RPP(*n*=1) heterostructures

The optical properties of another RPP flake that was partially covered by a large hBN flake is studied here by utilizing the fiber-collected CL spectroscopy. First, we scan the sample by moving the electron-beam excitation spot while keeping the position of the detecting fiber fixed (at the cross mark in Supplementary Figure 1a). Second, the detecting fiber is moved over the heterostructure, while keeping the electron excitation position fixed (at the cross mark in Supplementary Figure 1d). Thereby, we would be able to determine the effect of changing the excitation and detection positions in the investigated hBN/RPP heterostructure on the CL emission, in addition to probing the propagation characteristic of the excited excitons at the hBN/RPP heterostructure over long-range distances. In the first measurement, when the electron-beam impacts the tapered-like edge of the unencapsulated RPP flake, edge excitons are efficiently excited, leading to a strong and broad CL peak (shown by orange zones and spectrum in Supplementary Figure 1(a-c)). However, when the electron-beam impacts the surface of the unencapsulated RPP flake the excitonic peak intensity decreases and remains constant, until the electron-beam impacts the hBN edge on top of RPP flake. Scanning to the hBN/RPP heterostructure by passing through the hBN edge (shown by red zones and spectrum in Supplementary Figure 1(a-c)), the excitonic peak increases in consistency with the luminescent enhancement in Fig. 1d and Fig. 2c, d, e. Next, passing over the underlying RPP edge and scanning on the extruded part of hBN (shown by black zones and spectrum in Supplementary Figure 1(a-c)) the excitonic peak intensity is decreased, but remains at a constant value. Scanning the electron-beam over the wrinkles in the extruded hBN flake leads to a higher CL intensity at the excitonic peak, which can be attributed to exciting higher local defects density in hBN that improves the resonance response of the coupled exciton-defect system. Moreover, the emergence of the narrowed, red-shifted CL spectra in exciting the hBN/RPP heterostructure (red peak), and the extruded part of hBN (black peak) are reconfirmed in Supplementary Figure 1(c).

In the second step of this measurement, by scanning the fiber away from the RPP edge, while continuously measuring the emitted CL spectra, we observe the excitonic peak at $535\ \mathrm{nm}$ on the extruded part of the hBN flake over a long distance up to 100 μm (Supplementary Figure 1(d, e)). Moreover, a higher excitonic CL intensity is collected at hBN wrinkles, which is related to higher portion of excitonic radiative decay at defects, discontinuities and edges (Supplementary Figure 1(e)). This measurement confirms a spatial exponentially decaying behavior of the excitonic peak intensity over the long-range, wherein an exponential curve with decaying coefficient of $\gamma =$ 154.9 μm is fitted to the average measured CL intensities (Supplementary Figure 1(f)), comparable with the measurement in Fig. 4d.

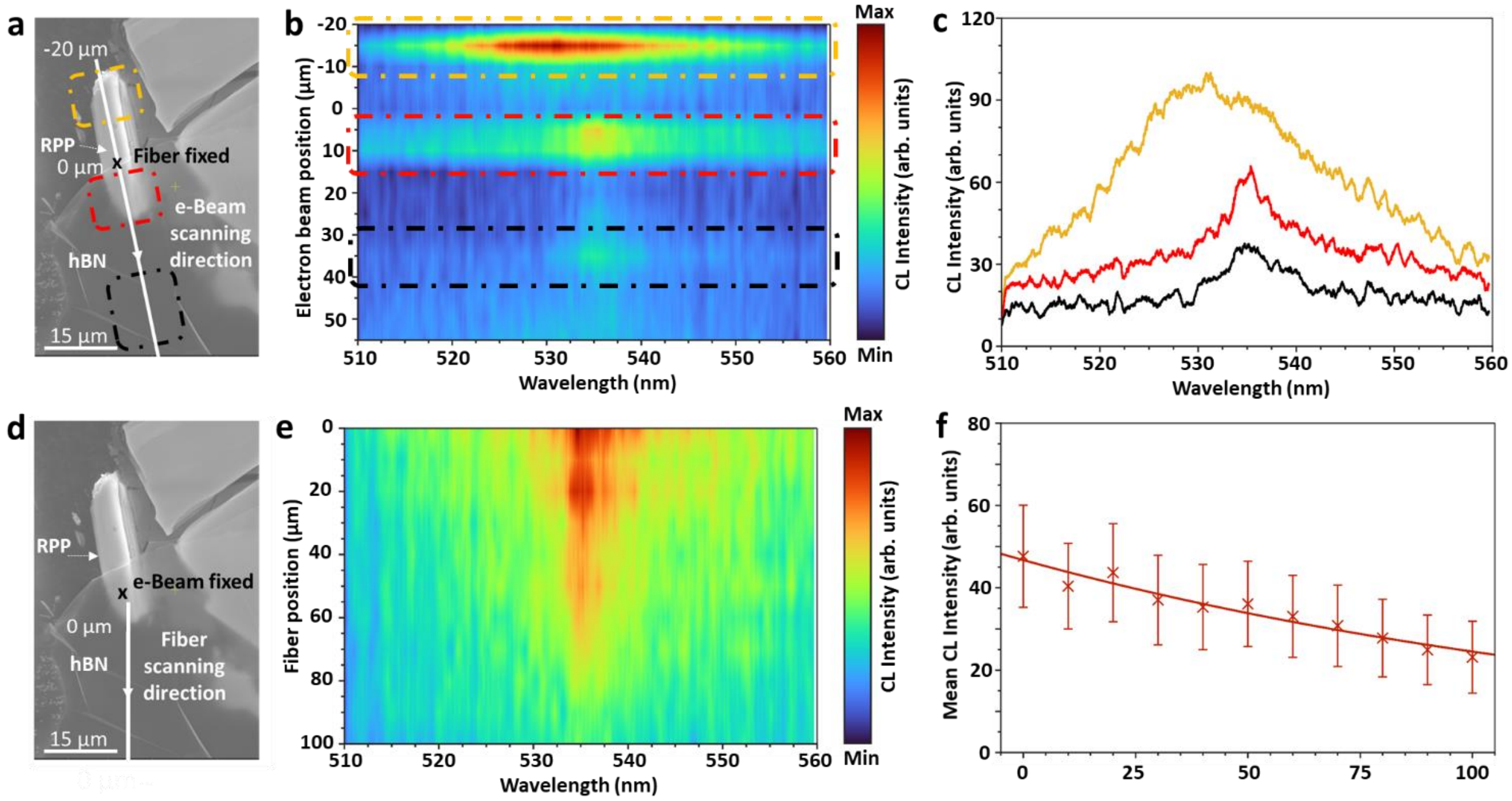

**Supplementary Figure 1: Fiber-collected CL spectroscopy results of the hBN/RPP heterostructure. Secondary electron SEM image of the investigated RPP structure is demonstrated in both (a) and (d).** (a) Fiber position is fixed on the sample at the cross mark, while electron-beam is scanning the structure along the white arrow. (b) The measured CL spectra along the electron-beam scanning path. The regions highlighted by the dash-dotted boxes correspond to the regions marked on the SEM image with the same colors. (c) CL spectra integrated over the selected regions specified by colored boxes in (b). (d) The electron-beam position is fixed on the hBN/RPP heterostructure at the cross mark, while the fiber is scanning along the white arrow. (e) The measured CL spectra along the fiber's scanning path in part (c). (f) The mean excitonic peak intensity and standard deviation of the measured CL spectra from part (e) versus the scanning fiber position, fitted to an exponential function of exp(-x/γ) with a decaying coefficient of γ=154.9 µm.

Spatial decaying behavior of the excitonic CL peak is also measured on a large pure RPP flake without any hBN flake. As illustrated in Supplementary Figure 2(a), electron-beam is fixed at the cross mark to excite RPP's excitons, and the CL-collecting fiber is scanning along the white arrow, up to 100 µm away from the excitation point. Supplementary Figure 2(b, c) exhibit the collected excitonic CL peak and the mean CL peak intensity versus the fiber position, revealing the fast-decaying behavior of the peak intensity in pure RPP along a few micrometers in contrast with the hBN/RPP heterostructure.

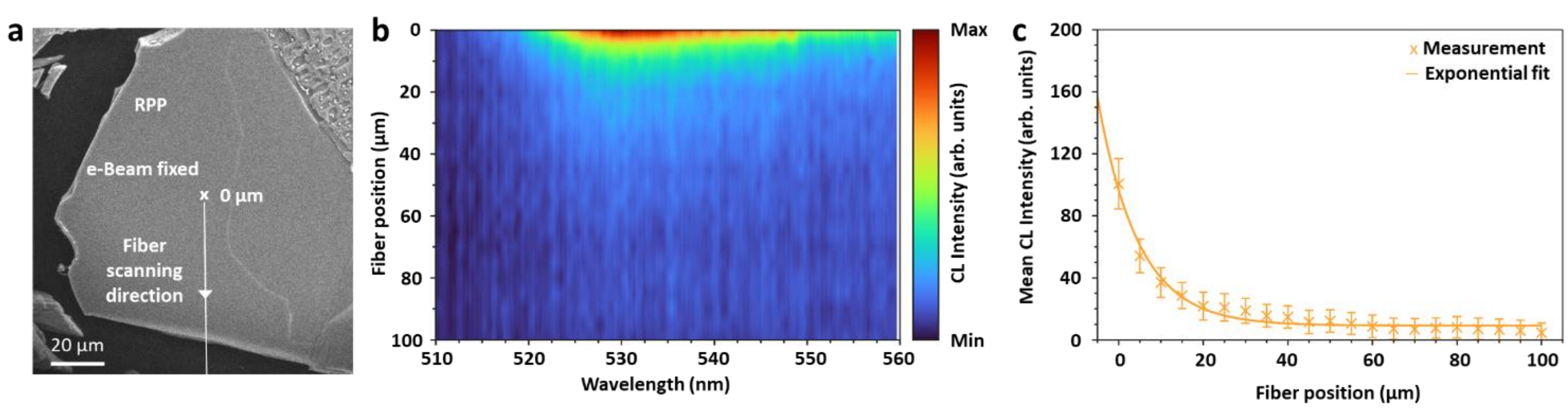

**Supplementary Figure 2 | Fiber-collected cathodoluminescence (CL) spectroscopy of a pristine RPP flake.** (a) Secondary electron SEM image of the investigated RPP flake. The electron-beam is fixed at the position marked by the cross, while the fiber is scanned along the direction indicated by the white arrow. (b) CL spectra measured along the fiber scanning path shown in (a). (c) Mean peak intensity (with standard deviation) of the CL spectra from (b) plotted as a function of fiber position. The data are fitted to an exponential decay function of the form $\exp(-x/\gamma)$, yielding a decay length of $\gamma = 9.6$ μm.

## Supplementary Note 2: Monte Carlo Simulations.

In order to investigate the contribution of the secondary electrons and backscattered electrons of individual layers to the CL response, we perform depth-resolved electron energy-loss simulations using CASINO Software[1-3]. This approach has been proven to efficiently reproduce CL responses beyond Eikonal approximations, when combined with Maxwell's equations[4]. In this package, the elastic scattering events have been calculated using the tabulated Mott cross-sections, and the inelastic interactions have been approximated by the mean energy loss rate between the successive collisions, as described by the modified Bethe law by Joy and Luo[1]. The simulations employ a single-scattering model to track the electron trajectories and compute energy loss, expressed as $\Delta E = \frac{dE}{dx}.d$, where $\frac{dE}{dx}$ is the mean energy loss rate, and $d$ is the distance between collisions[1,2]. Trajectory calculations terminate when electrons exit the sample, or their energy falls below a threshold value of 50 eV. For each scan point, the total deposited energy is calculated by summing energy loss contributions across the volume to generate energy-loss maps, revealing spatial distributions of energy absorption and scattering in the hBN/RPP heterostructures (Supplementary Figure 3a to c). The energy deposition can also be analyzed as linear scans or cross-sectional views, providing insights into dynamic or static beam-sample interactions (Supplementary Figure 3d to f).

The electron-beam is assumed to have a beam diameter of 10 nm, primary beam energy of 20 kV, and an incident electron number of 200,000. Cross-sectional views of energy-loss maps as a function of the electron-beam position on a hBN/RPP heterostructure show the amount of the absorbed energy within the sample, scaling from black (no absorbed energy) to white (maximum absorbed energy) in Supplementary Figure 3a to c. As the penetrating electrons interact with the layers, they undergo both elastic and inelastic scatterings, resulting in their lateral and vertical trajectories as well as the generation of secondary electrons. When the electron-beam impinges on the sample, 2000 nm away (on the left side) from the vacuum/RPP boundary, the electrons undergo significant energy loss in the top hBN layer and in the bottom carbon substrate (Supplementary Figure 3a).

For the case of electron-beam impinging directly at the RPP edge, the electrons that penetrate the hBN/RPP region experience significant inelastic scatterings due to the underlying RPP layer in the right half (Supplementary Figure 3b). When the electron-beam impinges 2000 nm away (right side) from the underlying vacuum/RPP boundary, a huge energy loss inside the RPP layer is observed (Supplementary Figure 3c). Better clarified by the linear scans, the maximum percentage of electron energy-loss occurs at a depth of approximately 3 μm in the bottom carbon tape when electron-beam impinges the extruded part of the hBN layer (Supplementary Figure 3d), whereas the majority of energy dissipation occurs within the RPP layer at a depth of approximately 1 µm for excitation on the hBN/RPP heterostructure (Supplementary Figure 3e). Moreover, a higher portion of electron energy loss occurs in the hBN layer when the underlying RPP presents below it in the hBN/RPP heterostructure, which is due to the scattered electrons from the RPP back into the hBN. As a clear confirmation, 2D simulation of the electron trajectories within the hBN/RPP heterostructure is presented in the inset of Supplementary Figure 3e, where red trajectories represent backscattered electrons and blue trajectories indicate absorbed electrons. Additionally, the fraction of secondary electrons captured in the RPP layer as a function of the electron-beam impinging x-position is plotted in Supplementary Figure 3f, wherein the electron-beam is scanned along x-axis over the RPP edge (at $x$=0 nm). As observed, the fraction of captured secondary electron in RPP that are responsible for the observed excitonic CL peak, increases with a sharp step-like transition at the RPP edge. Therefore, the simulated short-range trajectory of secondary electrons in the extruded part of the hBN layer (Supplementary Figure 3f) confirms that the observed CL signal, when the electron-beam impacts the structure at locations far from the RPP flake, cannot be attributed to the excitation of RPP excitons by backscattered electrons.

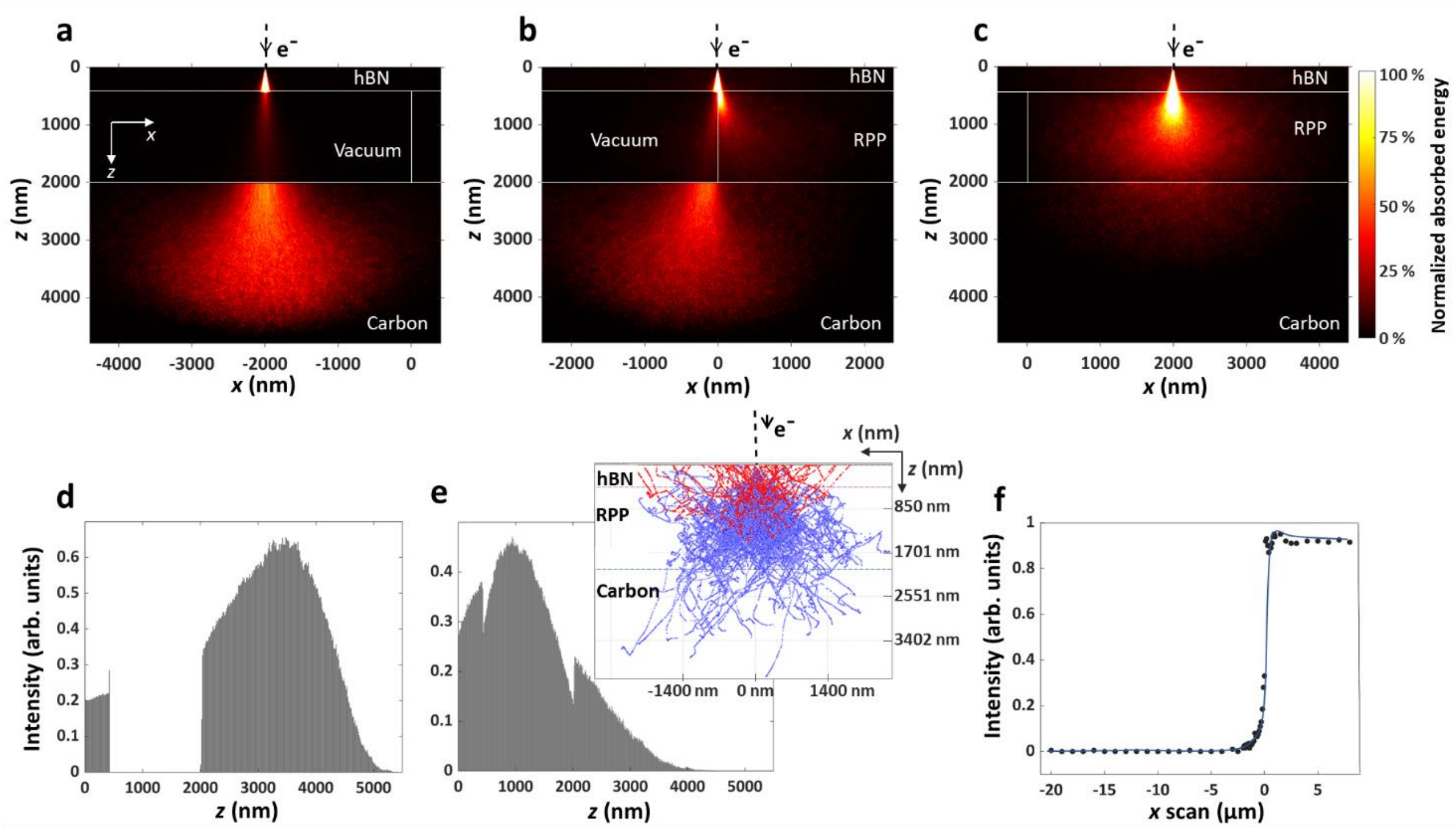

**Supplementary Figure 3: Monte Carlo simulations of the spatial distribution of inelastically scattered electrons in the hBN/RPP heterostructure.** Simulated energy-loss maps (480 × 480 points) are shown as a function of electron-beam scanning position on (a) a suspended hBN film over a carbon substrate at the distance of 2000 nm away from the RPP/vacuum interface, (b) the vacuum/RPP interface positioned on a carbon substrate, and (c) the hBN/RPP heterostructure at the distance of 2000 nm away from vacuum edge. Each map is generated with 200,000 incident electrons, and a beam diameter of 10 nm. Simulated electron energy-loss profiles for (d) the hBN/Vacuum/Carbon structure and (e) the hBN/RPP/Carbon structure. The inset illustrates electron trajectories of the secondary electrons (blue) and backscattered electrons (red)

generated by a 20 kV electron-beam. (f) Simulated secondary-electron fraction captured in RPP as a function of electron-beam position from the RPP/vacuum edge.

## Supplementary Note 3: The effect of RPP degradation on the optical response of the hBN/RPP heterostructure

The effect of RPP's inherent degradation on the CL characteristic of the hBN/RPP heterostructure is investigated by the mirror-collected CL spectroscopy technique. The secondary electron SEM image and the measured CL spectra of the freshly prepared hBN/RPP heterostructure are presented in Supplementary Figure 4a and b, and comparing it with the CL spectra of the same heterostructure after 30 days in Supplementary Figure 4c and d. The CL measurements are collected during moving the electron-beam excitation spot along the blue arrows in parts a and c, at low current regime of 200 pA and acceleration voltage of 20 kV. The degraded underlying RPP experiences from a more significant electron charging effect and a reduced quality in the achieved secondary electron SEM image in Supplementary Figure 4c, as compared with Supplementary Figure 4a. Comparing the measured CL spectra in Supplementary Figure 4b and d reveals that the enhanced peak at 525 nm on the fresh hBN/RPP heterostructure diminishes to a very weak trace of excitonic peak in the degraded sample. This evidence solidifies the proposed exciton-defect coupling in hBN/RPP heterostructure as the main origin of the observed enhanced CL at 525 nm before degradation of RPP, as opposed to other effects such as contribution of the secondary electrons and backscattered electrons of individual layers, such as the bottom RPP layer, to the CL response.

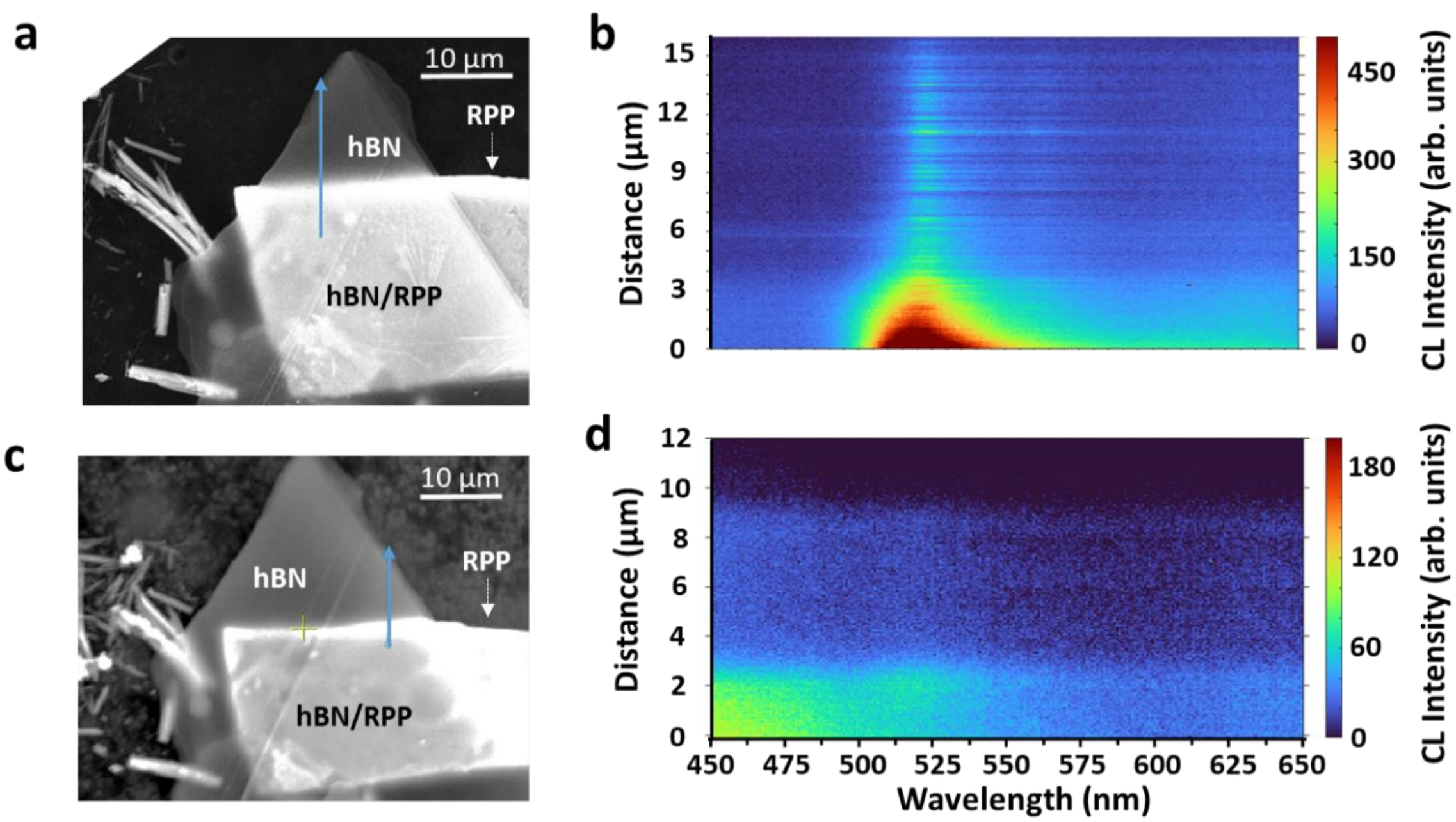

**Supplementary Figure 4: Effect of one-month aging in RPP explored by the mirror-collected CL system.** SEM image of (a) the prepared fresh hBN/RPP heterostructure, and (c) the same structure after one month, wherein blue arrows show the electron-beam scanning length. The measured CL spectra along the electron-beam scanning length (b) for the fresh hBN/RPP sample, and (d) for the same hBN/RPP sample after one month.

**Supplementary Note 4: Exploring the hBN/RPP(n=2) heterostructure by mirror-collecting cathodoluminescence.**

To elaborate more on the exciton-defect coupling in hBN/RPP heterostructure, hBN/RPP(*n*=2) heterostructures are realized and probed by the mirror-collected CL spectroscopy setup (Supplementary Figure 5). In this CL measurement, the electron-beam with 200pA current is moved along the blue arrow on the hBN/RPP(*n*=2) (part a), passing over the underlying RPP edge, and collecting the emitted photons. The measured energy-distance CL intensity maps and the individual CL spectra at different positions marked by '1' to '4' in part (a) reveal a dominant CL peak at 580 nm and a weak peak at 530 nm on the hBN/RPP(*n*=2) heterostructure (marked by '4'). Moreover, the dominant peak intensity is drastically decreased by sweeping the electron-beam toward the extruded part of the hBN away from the underlying edge of RPP (marked by '1'). However, the weak peak at 530 nm is still observable with a comparable intensity on the extruded hBN at position marked by '1'. To clarify the exciton-defect coupling in the hBN/RPP(*n*=2) heterostructure, the CL spectrum of a pure RPP(*n*=2) is superimposed on the CL trace of a pure hBN flake, acquired at a high beam current of 7 nA (part d). This experiment strongly proves the dominant role of exciton wavelength in emergence of the enhanced CL peak due to the coupled exciton-defect state at the hBN/RPP heterostructure. Here, excitons of RPPs(*n*=2) at 590 nm show significant spectral overlap with the phonon sidebands of hBN defects at 570 nm, and negligible spectral overlap with the defect peak at 530 nm (Supplementary Figure 5d), leading to emergence of a dominant peak at about 580 nm and a weak peak at around 530 nm on the hBN/RPP(*n*=2) heterostructure. Thus, by changing the exciton energy in the underlying RPP by changing the *n* value during the chemical synthesis step, we can tune the energy of the coupled exciton-defect states in the hBN/RPP heterostructures.

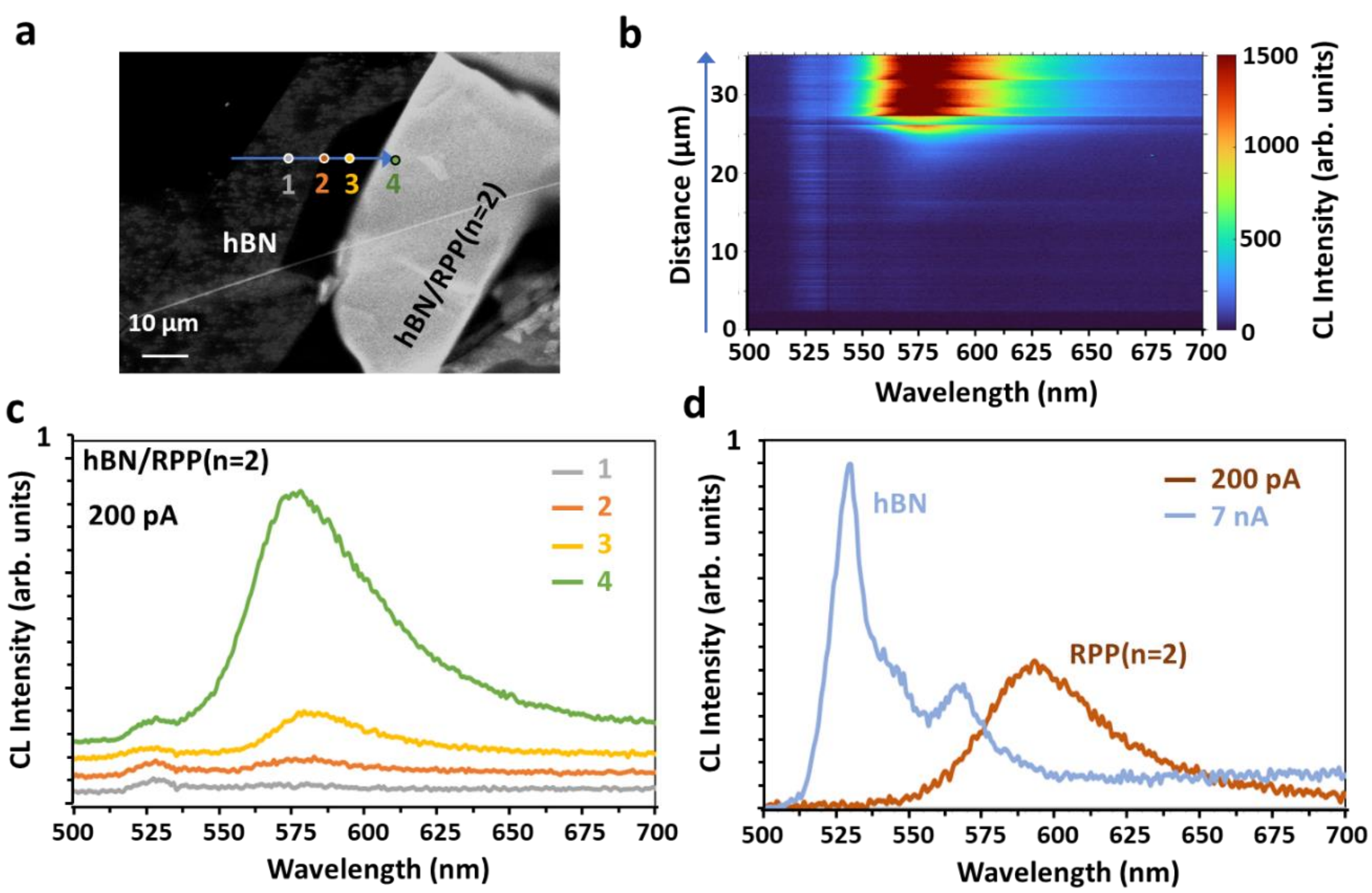

**Supplementary Figure 5: Spatially-resolved cathodoluminescence response of the hBN/RPP(*n*=2) heterostructure.** The exciton-defect coupling in hBN/RPP(*n*=2) has been studied the mirror-collected CL spectroscopy setup at the beam current of 200 pA and acceleration voltage of 20 kV (a) Secondary-electron SEM image of the hBN/RPP(*n*=2), wherein the blue arrow shows the electron-beam scanning path across the underlying RPP(*n*=2) edge. (b) The measured CL spectra along the blue arrow in part (a). (c) Individual CL spectra corresponding to four different spots, marked as '1', '2', '3' and '4' in part (a). CL spectra of a pure hBN flake (without RPP) integrated over a large area at the low current (200 pA) and high current (7 nA) regimes, superimposed on the CL spectrum of a pure RPP(*n*=2) flake at low current (200 pA) regime.

## Supplementary Note 5: Exploring the electron-beam-induced degradation in RPPs

To investigate the suitable conditions for CL measurements on RPP-based samples, we utilize two approaches: (i) benefiting from hBN-encapsulation in the hBN/RPP heterostructure, and (ii) finding the appropriate electron-beam conditions to achieve an efficient exciton excitation and avoiding the degradation issue in RPPs.

To evaluate the protection functionality of the hBN layer against electron-beam irradiation, RPP flakes that are partially covered by a large hBN flake (Supplementary Figure 6a) are exposed to an intense electron-beam condition at high sample current of 16 nA and acceleration voltage of 10 kV in the scanning electron microscope (SEM). The uncovered part of RPP flakes show a drastic electron-beam-induced degradation, while the covered part is protected (Supplementary Figure 6b). Therefore, the realized hBN/RPP heterostructure surpasses the electron-beam-induced degradation in CL measurements.

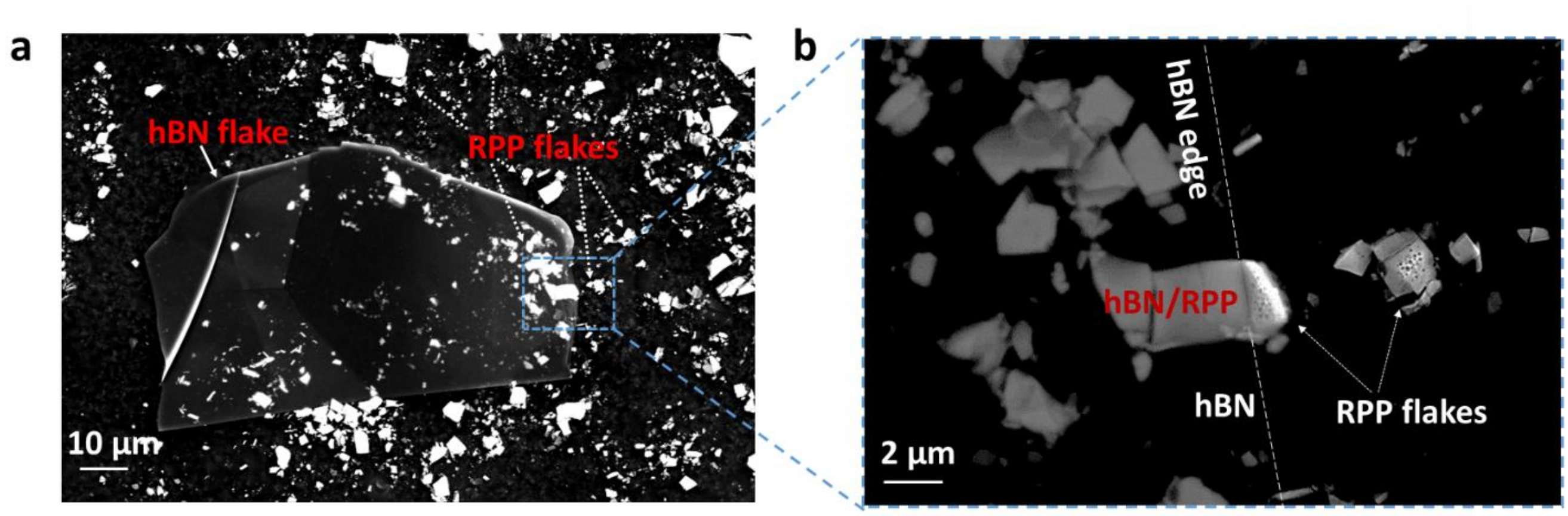

**Supplementary Figure 6: Protection efficiency of hBN encapsulation of RPP flakes against an intense electron-beam condition, at a high current of 16 nA, acceleration voltage of 10 kV and acquisition time of 500 ms in the SEM apparatus.** (a) Secondary-electron SEM image of a large hBN flake exfoliated on multiple RPP flakes. (b)The magnified view of the partially covered RPP flake at the edge of hBN flake. Uncovered parts of RPP flakes (right side) are drastically degraded, while encapsulation by hBN flake has significantly hindered the electron-beam-induced degradation.

To investigate the efficiency of exciton excitation by electron-beam in unencapsulated RPP flakes, while keeping a low degradation rate, we change the acceleration voltage (10 kV, 15 kV and 20 kV in Supplementary Figure 7) at a fixed low current of 200 pA in the mirror-collected CL spectroscopy. Efficient excitation of bulk excitons along the scanning length by the electron-beam (shown by blue arrows) is confirmed for acceleration voltage of 20 kV (Supplementary Figure 7a), as opposed to lower acceleration voltages that lead to excitation of excitons merely at the flake edge (Supplementary Figure 7b and 7d). Moreover, Supplementary Figure 7 does not show any detectable degradation trace either in the secondary electron images of RPP flakes, nor in the CL peak intensity when using low electron currents. Hence, we fix the acceleration voltage at 20 kV and the electron-beam current below 200 pA, where the current is measured at the sample position, to investigate RPPs and the hBN/RPP heterostructures by cathodoluminescence.

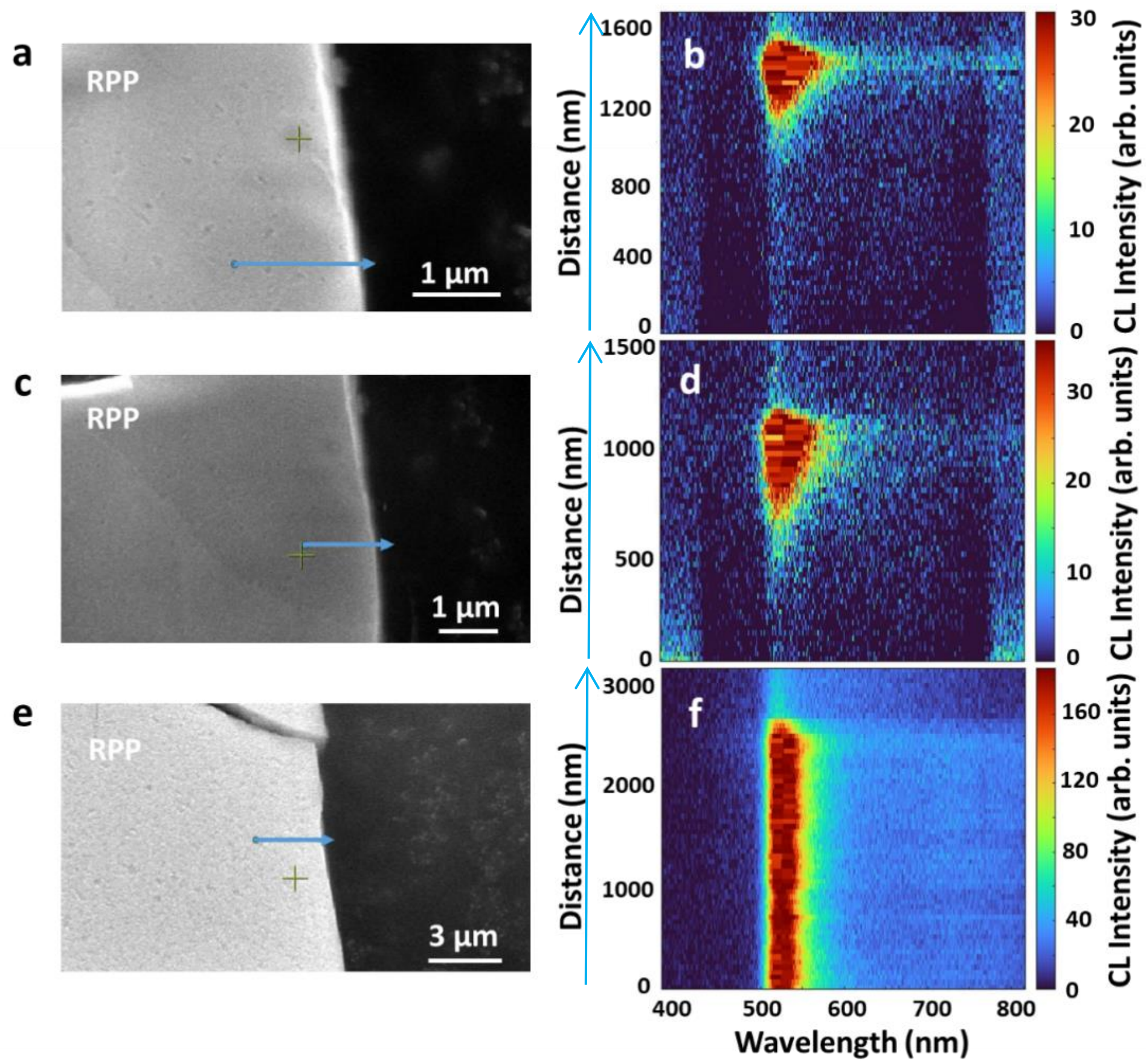

**Supplementary Figure 7: Exciton excitation efficiency of the unencapsulated RPP flakes at different acceleration voltages explored by the mirror-collected CL spectroscopy.** Scanning the electron-beam along the blue arrows shown in the left images in parts (a, c, e). Current level is fixed at 200 pA for all measurements to avoid the electron-beam-induced degradation in RPP. The measured CL spectra along the scanning length at acceleration voltages of (b) 10 kV, (d) 15 kV, and (f) 20 kV. Acceleration voltages below 20 kV leads to exciton excitation just at flake edges, while (f) proves efficient excitation of bulk excitons in RPP flakes without considerable degradation.